\documentclass[runningheads]{llncs}

\usepackage{algorithm}
\usepackage{graphicx}
\usepackage{subcaption}
\usepackage{multicol}
\usepackage{xcolor}
\graphicspath{ {./images/} }
\usepackage[noend]{algpseudocode}
\usepackage{multirow}

\usepackage[utf8]{inputenc}

\usepackage{algorithm}
\usepackage{algpseudocode}
\usepackage{amsmath}
\usepackage[breaklinks=true]{hyperref}

\begin{document}
	
	\title{\sc{$m$-Cubes}: An efficient and  portable implementation of Multi-Dimensional Integration for GPUs \thanks{ Work supported by the Fermi National Accelerator Laboratory, managed and operated
			by Fermi Research Alliance, LLC under Contract No. DE-AC02-07CH11359 with the U.S.
			Department of Energy. The U.S. Government retains and the publisher, by accepting the
			article for publication, acknowledges that the U.S. Government retains a non-exclusive,
			paid-up, irrevocable, world-wide license to publish or reproduce the published form of
			this manuscript, or allow others to do so, for U.S. Government purposes.
FERMILAB-CONF-22-043-LDRD-SCD \\ We acknowledge the support of Jefferson Lab grant to Old Dominion University 16-347. Authored by Jefferson Science Associates, LLC under U.S. DOE Contract No. DE-AC05-06OR23177 and DE-AC02- 06CH11357.\\ We thank Mahsa Sharifi for contributing to the initial implementation of $m$-Cubes. \\ code available at https://github.com/marcpaterno/gpuintegration}}
	
	\authorrunning{I. Sakiotis et al.}
	\titlerunning{\sc{$m$-Cubes}: An efficient and  portable implementation of Multi-Dimensional Integration for GPUs}
	
	\author{
		Ioannis Sakiotis\inst{1} \and
		Kamesh Arumugam\inst{2} \and
		Marc Paterno\inst{3}\and
		Desh Ranjan\inst{1}\and
		Bal{\v s}a Terzi\'c\inst{1}\and
		Mohammad Zubair\inst{1}
	}
	
	\authorrunning{Ioannis Sakiotis et al.}
	
	\institute{
		Old Dominion University, Norfolk, VA 23529, USA \and
		nvidia, Santa Clara, CA 95051-0952, USA \and
		Fermi National Accelerator Laboratory, Batavia, IL 60510
	}

	\maketitle 
	
	\begin{abstract}
		
	The task of multi-dimensional numerical integration is frequently encountered in physics and other scientific fields, e.g., in modeling the effects of systematic uncertainties in physical systems and in Bayesian parameter estimation. Multi-dimensional integration is often time-prohibitive on CPUs. Efficient implementation on many-core architectures is challenging as the workload across the integration space cannot be predicted a priori. We propose \textsc{$m$-Cubes}, a novel implementation of the well-known \textsc{Vegas} algorithm for execution on GPUs. \textsc{Vegas} transforms integration variables followed by calculation of a Monte Carlo integral estimate using adaptive partitioning of the resulting space. \textsc{$m$-Cubes} improves performance on GPUs by maintaining relatively uniform workload across the processors. As a result, our optimized \textsc{Cuda} implementation for \textsc{Nvidia} GPUs outperforms parallelization approaches proposed in past literature. We further demonstrate the efficiency of \textsc{$m$-Cubes} by evaluating a six-dimensional integral from a cosmology application, achieving significant speedup and greater precision than the \textsc{Cuba} library’s CPU implementation of \textsc{Vegas}. We also evaluate \textsc{$m$-Cubes} on a standard integrand test suite. Our approach yields a speedup of at least $10$ when compared against publicly available Monte Carlo based GPU implementations. In summary, \textsc{$m$-Cubes} can solve  integrals that are prohibitively expensive using standard libraries and custom implementations. A modern C++ interface header-only implementation makes \textsc{$m$-Cubes} portable, allowing its utilization in complicated pipelines with easy to define stateful integrals. Compatibility with non-\textsc{Nvidia} GPUs is achieved with our initial implementation of \textsc{$m$-Cubes} using the Kokkos framework. 
	\end{abstract}
	
	\section{Introduction}
	
	The task of multi-dimensional numerical integration is often encountered in physics and other scientific fields, e.g., in modeling the effects of systematic uncertainties in physical systems and Bayesian parameter estimation. However, multi-dimensional integration is time-prohibitive on CPUs. The emerging high-performance architectures that utilize accelerators such as GPUs can speed up the multi-dimensional integration computation. The GPU device is best suited for computations that can be executed concurrently on multiple data elements. In general, a computation is partitioned into thousands of fine-grained operations, which are assigned to thousands of threads on a GPU device for parallel execution. 

	A naive way to parallelize the multi-dimensional integration computation is as follows: divide the integration region into ``many'' ($m$) smaller sub-regions,  estimate the integral in each sub-region individually ($I_i$) and simply add these estimates to get an estimate for the integral over the entire region ($\Sigma_{i=1}^{m} I_{i}$). The integral estimate in each sub-region can be computed using any of the traditional techniques such as quadrature or Monte Carlo based algorithms. If we use a simple way of creating the sub-regions, e.g., via dividing each dimension into $g$ equal parts, the boundaries of the sub-regions are easy to calculate, and the estimation of the integral in different sub-regions can be carried out in an ``embarrassingly parallel'' fashion. Unfortunately, this approach is infeasible for higher dimensions as the number of sub-regions grows exponentially with the number of dimensions $d$. For example if $d = 10$ and we need to split each dimension into $g = 20$ parts the number of sub-regions created would be $g^d = 20^{10}$ which is roughly $10^{13}$. Moreover, uniform division of the integration region is not the best way to estimate the integral. The intuition is that the regions where the integrand is ``well-behaved'' do not need to be sub-divided finely to get a good estimate of the integral. Regions where it is ``ill-behaved'' (e.g. sharp peaks, many oscillations) require finer sub-division for a reliable, accurate estimate. However, when devising a general numerical integration method, we cannot assume knowledge of the behavior of the integrand. Hence, we cannot split up the integration region in advance with fewer (but perhaps larger in volume) sub-regions where the integrand is ``well-behaved'' and a greater number of smaller sub-regions in the region where it is ``ill-behaved'. To summarize, efficient implementation of multi-dimensional integration on many-core architectures such as GPUs is challenging due to two reasons: (i) increase in computational complexity as the dimension of the integration space increases, and (ii) the workload across the integration space cannot be predicted. 

	The first challenge, ``curse of dimensionality'', can be addressed to some extent by using a Monte Carlo based algorithm for multi-dimensional integration, as the convergence rate of such methods is independent of the dimension $d$. The convergence rate can sometimes be further improved by utilizing low-discrepancy sequences (Quasi-Monte Carlo) instead of pseudo-random samples \cite{qmcGPU} \cite{Goda2019RecentAI}. When utilizing Monte Carlo based approaches, the second challenge of consolidating sampling efforts on the ``ill-behaved' areas of the integration space, is addressed through ``stratified'' and/or ``importance'' sampling, which aim to reduce the variance of the random samples. Stratified sampling involves sampling from disjoint partitions of the integration space, the boundaries of which can be refined recursively in a manner similar to adaptive quadrature. Importance sampling integration methods, use Monte Carlo samples to approximate behavior of the integrand  in order to sample from a distribution which would significantly reduce the variance and accelerate convergence rates. This is accomplished by an initially uniform weight function that is refined across iterations, and results in more samples in the location where the magnitude of the integrand is either large or varies significantly. 

	The sequential \textsc{Vegas} algorithm is the most popular Monte Carlo method that makes use of importance sampling \cite{LEPAGE2021110386} \cite{PETERLEPAGE1978192}. There are several implementations and variants, including Python packages, C++-based implementations in the \textsc{Cuba} and \textsc{Gsl} libraries, and the R Cubature package. Unfortunately, while \textsc{Vegas} can often outperform standard Monte Carlo and deterministic techniques, sequential execution often leads to prohibitively long computation times. A GPU implementation of the original \textsc{Vegas} algorithm was proposed in \cite{KanzakiJ2011MCio}, but is not packaged as a library though an implementation exists in \cite{madgraph}. VegasFlow is a Python library based on the TensorFlow framework, providing access to \textsc{Vegas} and standard Monte Carlo implementations that can execute on both single and multi-GPU systems \cite{Carrazza} \cite{vegasflow_package}. Another Python package with support for GPU execution was proposed in \cite{WU2020106962}, incorporating stratified sampling and a heuristic tree search algorithm. All GPU implementations demonstrate significant speedup over serial versions of \textsc{Vegas} but impose restrictions on the required computing platforms and programming environments, e.g., the \textsc{Cuda} implementation of \cite{KanzakiJ2011MCio} requires an \textsc{Nvidia} GPU.

	We propose \textsc{$m$-Cubes}, a novel implementation of the well-known \textsc{Vegas} algorithm for multi-dimensional integration on GPUs. \textsc{$m$-Cubes} exploits parallelism afforded by GPUs in a way that avoids the potential non-uniform distribution of workload and makes near-optimal use of the hardware resources. Our implementation also modifies \textsc{Vegas} to make the computation even faster for functions that are ``fully symmetric".  Our approach demonstrates significant performance improvement over \cite{WU2020106962} and \cite{madgraph}. Our initial implementation was targeted for \textsc{Nvidia} GPUs utilizing \textsc{Cuda}, but the \textsc{$m$-Cubes} algorithm is applicable to any many-core system. Utilization of the Kokkos programming model allows execution on various parallel platforms including non-\textsc{Nvidia} GPUs and even CPU-clusters.
	Our goal is to make publicly available a robust, portable and easy-to-use, implementation of \textsc{Vegas} in \textsc{Cuda} and Kokkos that will be suitable for the execution of challenging integrands that occur in physics and other scientific fields. 

	The remainder of the paper is structured as follows. In section II, we describe various Monte Carlo based algorithms. In section III, we describe the \textsc{Vegas} algorithm. In section IV we describe the \textsc{$m$-Cubes} algorithm. In section V, we discuss the accuracy and performance of our implementation, comparing its execution time against publicly available Monte Carlo based methods. In section VI we discuss the interface and portability features used on an complex integral utilized in parameter estimation in cosmological models of galaxy clusters. Section VII presents results on an initial Kokkos implementation.

	\section{Background}
	We summarize here the previous work related to our research. We first summarize the previously developed sequential Monte Carlo Methods and libraries. Thereafter we summarize the research on parallel \textsc{Vegas} based methods. 
	
	\subsection{Monte Carlo Methods}
	The \textsc{Gsl} library provides three Monte Carlo based methods, standard Monte Carlo, \textsc{Miser}, and \textsc{Vegas}. Standard Monte Carlo iteratively samples the integration space of volume $V$, at $T$ random points $x_i$ to generate an integral estimate in the form of $\frac{V}{T}\sum_{t=i}^{T} f(x_i)$, whose error-estimate is represented by the standard deviation.
	
	\textsc{Vegas} is an iterative Monte Carlo based method that utilizes stratified sampling along with importance sampling to reduce standard Monte Carlo variance and accelerate convergence rates. The stratified sampling is done by partitioning the $d$-dimensional space into sub-cubes and computing Monte Carlo estimates in each  sub-cube. For importance sampling, \textsc{Vegas} samples from a probability distribution that is progressively refined among iterations to approximate the target-integrand. \textsc{Vegas} uses a piece-wise weight function to model the probability distribution, where the weight corresponds to the magnitude of the integral contribution of each particular partition in the integration space. At each iteration \textsc{Vegas} adjusts the weights and the corresponding boundaries in the integration space, based on a histogram of weights. The piece-wise weight-function is intentionally separable to keep the number of required bins small even on high-dimensional integrands. Existing implementations of \textsc{Vegas}, are also found within the \textsc{Cuba} and \textsc{Gsl} libraries. A Python package also exists, with support for parallelization through multiple processors. 

	\textsc{Miser} is another Monte Carlo based method, which utilizes recursive stratified sampling until reaching a user-specified recursion-depth, at which point standard Monte Carlo is used on each of the generated sub-regions. \textsc{Miser} generates sub-regions by sub-dividing regions on a single coordinate-axis and redistributing the number of sampling points dedicated to each partition in order to minimize their combined variance. The variance in each sub-region  is estimated at each step with a small fraction of the total points per step. The axis to split for each sub-region, is determined based on which partition/point-redistribution will yield the smallest combined variance.
	
	\textsc{Cuba} is another library that provides numerous Monte Carlo based methods (\textsc{Vegas}, Suave, Divonne). Suave utilizes importance sampling similar to \textsc{Vegas} but further utilizes recursive sub-division of the sub-regions like \textsc{Miser} in \textsc{Gsl}. The algorithm first samples the integration space based on a separable weight function (mirroring \textsc{Vegas}) and then partitions the integration space in two similar to \textsc{Miser}. Suave then selects the sub-region with the highest error for further sampling and partitioning. This method requires more memory than both \textsc{Vegas} and \textsc{Miser}.
	
	Divonne uses stratified sampling, attempting to partition regions such that they have equal difference between their maximum and minimum integrand values. It utilizes numerical optimization techniques to find those minimum/maximum values. Divonne can be faster than \textsc{Vegas}/Suave on many integrands while also providing non-statistically based error-estimates if quadrature rules are used instead of random samples. 
	
	\subsection{Parallel Programming Models}
	
	\textsc{Cuda} is a popular low-level programming model, allowing the execution of parallel computations on \textsc{Nvidia} GPUs and has been used extensively in scientific computing. The \textsc{Nvidia} GPU restriction can be avoided by utilizing Kokkos, a C++ library that implements an abstract thread parallel programming model which enables writing performance portable applications for major multi- and many-core HPC platforms. It exposes a single programming interface and allows the use of different optimizations for backends such as \textsc{Cuda}, HIP, SYCL, HPX, OpenMP, and C++ threads \cite{9485033}, \cite{CarterEdwards20143202}. Parameter tuning in the parallel dispatch of code-segments, allows for both automatic and manual adjustments in order to exploit certain architectural features.
	
	\subsection{Parallel GPU Methods}
	
	The \textsc{$g$Vegas}  method is a \textsc{Cuda} implementation of \textsc{Vegas} that that allows execution on a GPU \cite{KanzakiJ2011MCio} \cite{madgraph}. This method parallelizes the computation over the sub-cubes used in \textsc{Vegas} for stratification. It uses an equal number of samples in each sub-cube as proposed in the original \textsc{Vegas} algorithm. It assigns a single thread to process each sub-cube, which is not very efficient and is discussed in Section IV. The importance sampling that requires keeping track of integral values in each bin (explained in the next section) is done on the CPU which slows down the overall computation. Additionally, the number of possible samples is limited due to their allocation on GPU memory which imposes performance limitations. These design choices are a product of their time as the implementation was developed in the early stages of the CUDA platform. A modernized version exists in \cite{gvegascp} but does not meet the statistical requirements related to the returned ${\chi^2}$ as indicated in \cite{LEPAGE2021110386}. For that reason, Section V includes comparison with the implementation of \cite{madgraph}.
	
	Non-C++ implementations are available as well. ZMCintegral is a Python library for performing Monte Carlo based multi-dimensional numerical integration on GPU platforms, with support for multiple-GPUs. The algorithm uses stratified sampling in addition to a heuristic tree search that applies Monte Carlo computations on different partitions of the integration space \cite{WU2020106962}.

	\section{The \sc{Vegas} Algorithm}

	\textsc{Vegas} is one of the most popular Monte Carlo based methods. It is an iterative algorithm, that attempts to approximate where the integrand varies the most with a separable function that is refined across iterations. The main steps of a \textsc{Vegas} iteration are listed in Algorithm 1. The input consists of an integrand $f$ which is of some dimensionality $d$, the number of bins $n_b$ on each dimensional axis, the number of samples $p$ per sub-cube, the bin boundaries stored in an $d$-dimensional list $B$, and the $d$-dimensional list $C$ which contains the contributions of each bin to the cumulative integral estimate. 
	
	Initially the integration space is sub-divided to a number of $d$-dimensional hyper-cubes, which we refer to as sub-cubes. \textsc{Vegas} processes each sub-cube independently with a for-loop at line 2. At each sub-cube, the algorithm generates an equal number of samples\footnote{Here, we focus on the original \textsc{Vegas} algorithm which uses equal number of samples in each sub-cube. The later versions of the algorithm deploy adaptive stratification that adjust the number of integral estimates used in each sub-cube. }, which are processed through the for-loop at line 3. To process a sample, \textsc{Vegas} generates $d$ random numbers in the range $(0,1)$ at line 4, corresponding to one point per dimensional-axis. Then at line 5, we transform the point $y$ from the domain of the unit hyper-cube $(0,1)$ to actual coordinates in the integration space. At line 6, we evaluate the integrand $f$ at the transformed point $x$, yielding the value $v$ which contributes to the cumulative integral estimate. Before proceeding to the next sample, we identify  at line 7 the bins that encompass each of the $d$ coordinates in $x$. We use the indices of those bins ($b[1:d]$) to increment their contribution ($v$) to the integral at line 8. Once the samples from all sub-cubes have been processed, we exit the twice-nested for-loop. At line 9, we use the bin contributions stored in $d$, to adjust the bin boundaries $B$ in such a way that bins of large contributions are smaller. This approach results in many small bins in the areas of the integration space where the integrand is largest or varies significantly, resulting in more samples being generated from those highly contributing bins. Finally, the contribution of each sample must be accumulated at line 10, to produce the Monte Carlo integral estimate and to compute the variance for the iteration. 
	
	The most desirable features of \textsc{Vegas} are its ``importance sampling'' which occurs by maintaining bin contributions and adjusting the bin boundaries at the end of each iteration. The use of sub-cubes introduces ``stratified sampling'' which can further reduce the variance of the Monte Carlo samples. Those two variance reduction techniques make \textsc{Vegas} successful in many practical cases and the independence of the sub-cubes and samples make the algorithm extremely parallelizable.
	
	\begin{algorithm}
		\caption{\textsc{Vegas}}
		\begin{algorithmic}[1]  
			\Procedure{\textsc{Vegas}}{$f$, $d$, $n_b$, $p$, $B$, $C$}  
			\Comment{Each iteration consists of the steps below}
			\For{all sub-cubes} 
				\For{ $i \gets 1$ to $p$} 
				\Comment{$f$ is evaluated at $p$ points in each sub-cube}
					\State $y_1,y_2, ..., y_d \gets$ generate $d$ points in range $(0,1)$ uniformly at random
					\State $x_1,x_2, ..., x_d \gets$ map vector $y$ to vector $x$ 
					\Comment{$f$ is evaluated at $x$} 
					\State $v \gets f(x_1,x_2, ..., x_d)$ 
					\State let $b_i$ denote the index of the bin to which $x_i$ belongs in dimension $i$ 
					\State increment $C[1][b_1]$, $C[2][b_2]$, ..., $C[d][b_d]$ by $v$ \Comment{Store bin contributions}
				\EndFor 
			\EndFor 
			\State $B[1:d][1:n_b] \gets$ adjust all bin boundaries based on $C[1:d][1:n_b]$
			\State $I,E \gets$ compute integral estimate/variance by accumulating $v$ \\
			\Return $I$, $E$
			\EndProcedure
		\end{algorithmic}
	\end{algorithm}
	
	\section{The Algorithm \sc{$m$-Cubes}}
	
	The main challenges of parallel numerical integrators are the ``curse of dimensionality'' and workload imbalances along the integration space. While high-dimensionality is made manageable by the use of the Monte Carlo estimate in \textsc{Vegas} (Algorithm 1), workload imbalances need to be addressed. This is particularly true for newer variations of the \textsc{Vegas} algorithm, which involve a non-uniform number of samples per sub-cube. Parallelization of \textsc{Vegas} poses additional challenges from the need to accumulate the results of multiple samples from different processors. In Algorithm 1, line 10 involves such an accumulation which requires processor synchronization. Furthermore, a race condition can occur at line 8, where the contributions of a bin may need to be updated by different processors.
	
	To parallelize \textsc{Vegas}, \textsc{$m$-Cubes} (Algorithm 2) assigns a batch of sub-cubes to each processor and generates a uniform number of samples per sub-cube. This solves the work-load imbalance issue and further limits the cost of accumulating results from various processors. The integrand contributions from all sub-cubes of each processor (Algorithm 1, line 6), are processed serially. As a result, those values can be accumulated in a single local variable, instead of synchronizing and transferring among processors. This does not eliminate the cost of accumulation, as we still need to collect the contributions from the sub-cube batches in each processor at line 10, but the extent of the required synchronization is reduced significantly.
		
	\begin{algorithm}
		\caption{\textsc{$m$-Cubes}}
		\begin{algorithmic}[1]
			
			\Procedure{$m$-Cubes}{$f, d, n_b, maxcalls, L, H, itmax, ita, r$}
			
				\State $I$, $E \gets 0$ \Comment{Integral/Error estimate}	
				\State $g \gets (maxcalls/2)^{{1}/{d}}$ \Comment{Number of intervals per axis}
				\State $m \gets {g}^d$	\Comment{Number of cubes}
				\State $s \gets$ \textsc{Set-Batch-Size}($maxcalls$) \Comment{Heuristic}
				\State $B[1:d][1:n_b] \gets $ \textsc{Init-Bins}($d$, $n_b$) \Comment{Initialize bin boundaries}
				\State $C[1:d][n_b] \gets 0$ \Comment{Bin contributions} 
				\State $p \gets {maxcalls}/{m}$	\Comment{number of samples per cube}
			
				\For{$i \gets 0$ to $ita$}
					\State $r$, $C \gets$ \textsc{V-Sample}()
					\State $I$, $E \gets$ \textsc{Weighted-Estimates}($r$)				
					\State $B \gets$ \textsc{Adjust-Bin-Bounds}($B, C$)
					\State \textsc{Check-Convergence}()
				\EndFor
			
				\For{$i \gets ita$ to $itmax$}			
					\State $r \gets$ \textsc{V-Sample-No-Adjust}()
					\State $I$, $E \gets$ \textsc{Weighted-Estimate}($r$)				
					\State \textsc{Check-Convergence}()
				\EndFor
			\EndProcedure
		\end{algorithmic}
	\end{algorithm}
	
	The input of the \textsc{$m$-Cubes} algorithm consists of the integrand $f$ and its dimensionality $d$, the number of bins per coordinate axis $n_b$, the maximum number of allowed integrand evaluations $maxcalls$, and the upper/lower boundaries of the integration space in each dimension, represented in the form of two arrays $L, H$. The user must also supply the required number of iterations $itmax$ and the number of iterations that will involve bin adjustments ($ita)$. We also use the array $r$ to store the results which consist of the integral estimate and standard deviation.
	
	In line 2, we initialize the cumulative integral estimate and error-estimate (standard deviation) to zero. In line 3 we compute the number of intervals per axis; the boundaries of the resulting sub-cubes remain constant for the duration of the algorithm. In contrast, the bin boundaries $B$ are adjusted across iterations. At line 4 we determine the number of sub-cubes $m$, while we also compute the batch size $s$, referring to the number of sub-cubes that each thread will process iteratively. Then the bin boundaries are generated on line 6, by equally partitioning each axis into $n_b$ bins, and storing their right boundaries in the list $B$. 
	
	Then we proceed with the \textsc{$m$-Cubes} iterations. The first step is to compute the result $r$ and bin contributions $C$, by executing the \textsc{V-Sample} method (Algorithm 3) at line 10. \textsc{V-Sample} produces the Monte Carlo samples, evaluates the integrals and updates the bin contributions. This method requires almost all data-structures and variables as parameters, so we omit them in this description. At line 11, the estimates are weighted by standard \textsc{Vegas} formulas that can be found in equations 5 and 6 of \cite{PETERLEPAGE1978192}. We then adjust the bin boundaries $B$ based on the bin contributions $C$. If the weighted integral estimate and standard deviation produced at line 11, satisfy the user's accuracy requirements, execution stops, otherwise we proceed to the next iteration. Before proceeding to the next iteration, the bin boundaries $B$ are adjusted at line 12. The only difference between an \textsc{$m$-Cubes} and a \textsc{Vegas} iteration from the original algorithm, are the parallelized accumulation steps and mappings between processors and sub-cubes. 
	
	A second loop of iterations (lines 14 to 17) is invoked once $ita$ iterations are completed. In this set of iterations, we perform the same computations with the exception of bin adjustments and their supporting computations which are omitted. This distinction is introduced due to the common occurrence of the boundaries $B$ converging after a number of iterations and remaining unchanged. In those cases, the costly operations of keeping track of bin contributions and updating them has no positive effect. As such, the user can mandate a limit of iterations with that will involve bin adjustments, and sub-subsequent iterations will execute faster by avoiding redundant operations.
	
	\textsc{V-Sample} and \textsc{V-Sample-No-Adjust} are the only methods that involve parallelization, encompassing the functionality of lines 2 to 8 from Algorithm 1. To facilitate the accumulation steps needed to yield the integral contributions from multiple sub-cube batches, \textsc{V-Sample} utilizes hierarchical parallelism, where each processor launches many groups of cooperative threads (\textsc{Cuda} thread-blocks/Kokkos teams) of size $x$, requiring a total $\frac{m}{x}$ such  groups. Each thread within a group is independent and processes its own sub-cube batch of size $s$ (see Algorithm 2, line 5). The benefit of this approach, is that  group-shared memory and group-synchronization capabilities allow for more efficient accumulation of the integral estimates $v$ local to each thread. The race condition involved with incrementing the bin contributions from multiple threads, is solved through atomic addition operations. The same operation is used to accumulate the integral estimate from all groups.
	
	The input of the \textsc{V-Sample} method, consists of the integrand $f$ of dimensionality $d$, the number of sub-cubes $m$, sub-cube batch size $s$, number of samples per sub-cube $p$, bin bounds $B$, bin contributions $C$, and result $r$. Once finished, \textsc{V-Sample} will return an estimate for the integral, variance, and updated bin contributions $C$.
	
	The for-loop at line 2, indicates the sequential processing of $s$ sub-cubes from each thread. At line 3 we initialize a random number generator. Each thread has local integral and error estimates $I$ and $E$ (line 4) respectively, which encompass the contributions from all $s$ sub-cubes. Each thread processes its assigned sub-cubes with the serial for-loop at line 5. As the sub-cubes are processed, the local estimates $I_t$ and $E_t$ of each sub-cube are accumulated in $I$ and $E$. This involves yet another for-loop at line 7, to serialize the $p$ samples generated per sub-cube. Similar to the accumulation of $I_t$ to $I$, we accumulate the estimates $I_k$ and $E_k$ (local to the sample) to $I_t$ and $E_t$.
	
	For each sample, we generate an $d$-dimensional point $x$ where we will evaluate the integrand $f$. This yields estimates for the sample that are used to increment the sub-cubes estimates at lines 10 and 11. Then, based on the bin IDs that are determined in line 12. we update the bin contributions in line 14. The atomic addition guarantees serial access for each thread updating $C$ at each index $b[1:s]$, avoiding race conditions. The actual bin-contribution is the square of the integral estimate $I_k$. Then, we update the variance at line 16, followed by the updating of the thread-local estimates for the entire batch of sub-cubes in lines 16 and 17. 
	
	Once the for-loop at line 5 is finished, we accumulate the $I$, $E$ from each thread in parallel. This is accomplished by a group-reduction that utilizes shared memory and warp-level primitives if available. Finally, once each group has accumulated estimates from all its sub-cubes across all its threads, a final atomic addition in lines 23 and 24 accumulates the estimates from all groups and can return them as the result $r$.
	
	The \textsc{V-Sample-No-Adjust} method is almost identical to \textsc{V-Sample}, with the distinction that the loop at lines 13-14 are not needed which yields a boost in performance. 
	
	\begin{algorithm}
		\caption{\textsc{V-Sample}}
		\begin{algorithmic}[1]	
			
			\Procedure{V-Sample}{$f, d, m, s, p, B, C, r$}
			
			\For{$m/b$ threads parallel}
			
				\State \textsc{Set-Random-Generator}($seed$)
				\State $I$, $E \gets 0$ \Comment{cumulative estimates of thread}

				\For{$t = 0$ to $s$}
				
					\State $I_t$, $ E_t \gets 0$	\Comment{estimates of sub-cube t}
					
					\For{$k \gets 1$ to $p$}
						\State $x[1:d] \gets$ \textsc{Generate}()
						\State $I_k, E_k \gets $ \textsc{Evaluate}($f,x$)
						\State $I_t \gets I_t + I_k$	\Comment{Accumulate sub-cube contributions}
						\State $E_t \gets E_t + E_k$
						
						\State $b[1:d] \gets$ \textsc{Get-Bin-ID}($x$)
						\For{$j \gets 1$ to $d$}	\Comment{Store bin contributions}
							\State \textsc{AtomicAdd}($C[b[j]]$, $I_k^2$)
						\EndFor  
			
					\EndFor	
					
					\State $E_t \gets$ \textsc{UpdateVariance}($E_t$, $I_t$, $p$)
					\State $I \gets I + I_t$ \Comment{update cumulative values}
					\State $E \gets E + E_t$
					
				\EndFor
				
				\State $I \gets$ \textsc{Reduce}($I$)
				\State $E \gets$ \textsc{Reduce}($E$)
				
				\If{thread 0 within group}
					\State \textsc{AtomicAdd}($r[0]$, $I$)
					\State \textsc{AtomicAdd}($r[1]$, $E$)
				\EndIf
				
			\EndFor 
			\EndProcedure
		\end{algorithmic}
	\end{algorithm}
	
	\section{Experimental Results}
	
	We performed two separate series of experiments to compare against the GPU methods gVegas and \textsc{ZMCintegral}. Our experiments utilized a standard integrand test suite (eq. $1$ to $6$) which consists of several integrals with different characteristics such as corner/product peaks, Gaussian, $C^0$  form, and oscillations. We used a single node with a $2.4$ GHz Xeon R Gold $6130$ CPU, v$100$ GPU with $16$GB of memory and $7.834$ Tflops in double precision floating point arithmetic, and compiled with \textsc{Gcc} 9.3.1 and \textsc{Cuda} $11$.
	 
	\begin{equation}
		f_{1,d}\left(x\right) = \cos\left(\sum_{i=1}^{d} i \, x_i\right)
	\end{equation}
	
	\begin{equation}
		f_{2,d}\left(x\right) = \prod_{i=1}^{d} \left(\frac{1}{50^2} + \left(x_i - 1/2\right)^2\right)^{-1}
	\end{equation}
	
	\begin{equation}
		f_{3,d}\left(x\right) = \left(1+ \sum_{i=1}^{d} i \, x_i\right)^{-d-1} 
	\end{equation}
	
	\begin{equation}
		f_{4,d}\left(x\right) = \exp\left(-625 \sum_{i=1}^{d} \left(x_i-1/2\right)^2\right)
	\end{equation}
	
	\begin{equation}
		f_{5,d}\left(x\right) =	\exp\left(-10  \sum_{i=1}^{d} | x_i - 1/2 |\right)
	\end{equation}
	
	\begin{equation}
		f_{6,d}\left(x\right) =	
			\begin{cases}
				\exp\left(\sum_{i=1}^d \left(i+4\right) x_i\right) & \text{ if } x_i < \left(3+i\right)/10 \\ 
			0 & \text{otherwise}
			\end{cases}
	\end{equation}
	
	\subsection{Accuracy}
		
	In the \textsc{$m$-Cubes} algorithm, we use relative error as a stopping criteria for achieving a specified accuracy, which is the normalized standard deviation (see Algorithm 3 for the error computation). The required accuracy associated with numerical integration, can vary significantly across applications depending on the context. To our knowledge no numerical integration algorithm can claim a zero absolute error on all integrands or even guarantee integral/error estimates that satisfy the various relative error tolerances $\tau_{rel}$. As such, it is important to evaluate the degree of correctness for specific challenging integrands whose integral values are known {\it a priori}. It is equally important to demonstrate how an algorithm adapts to increasingly more demanding precision requirements and whether the yielded integral/error estimates truly satisfy the user's required $\tau_{rel}$. This is especially true for Monte Carlo based algorithms, whose randomness and statistically-based error-estimates make them less robust than deterministic, quadrature-based algorithms. In our evaluation of \textsc{$m$-Cubes}, we adopt the testing procedures of \cite{GenzTest} in selecting the target integrands but preselect the various integrand parameter constants as in \cite{pagani}. We deviate from \cite{pagani}, in that we omit the two box-integrands that were not challenging for \textsc{Vegas}. We also do not report results on $f_{1,d}$ in our plots, as no \textsc{Vegas} variant could evaluate it to the satisfactory precision levels. The various tolerances are the same on all experiments as each integrand is evaluated on increasingly smaller $\tau_{rel}$.  We start evaluating all integrands with a $\tau_{rel}$ of $10^{-3}$. Upon each successful execution, we divide $\tau_{rel}$ by five until either surpassing the minimum value of $10^{-9}$ (maximum accuracy) or failing to achieve the required error tolerance.
	
\begin{figure}[hbt]
		\centering
		\includegraphics[width=.95\textwidth]{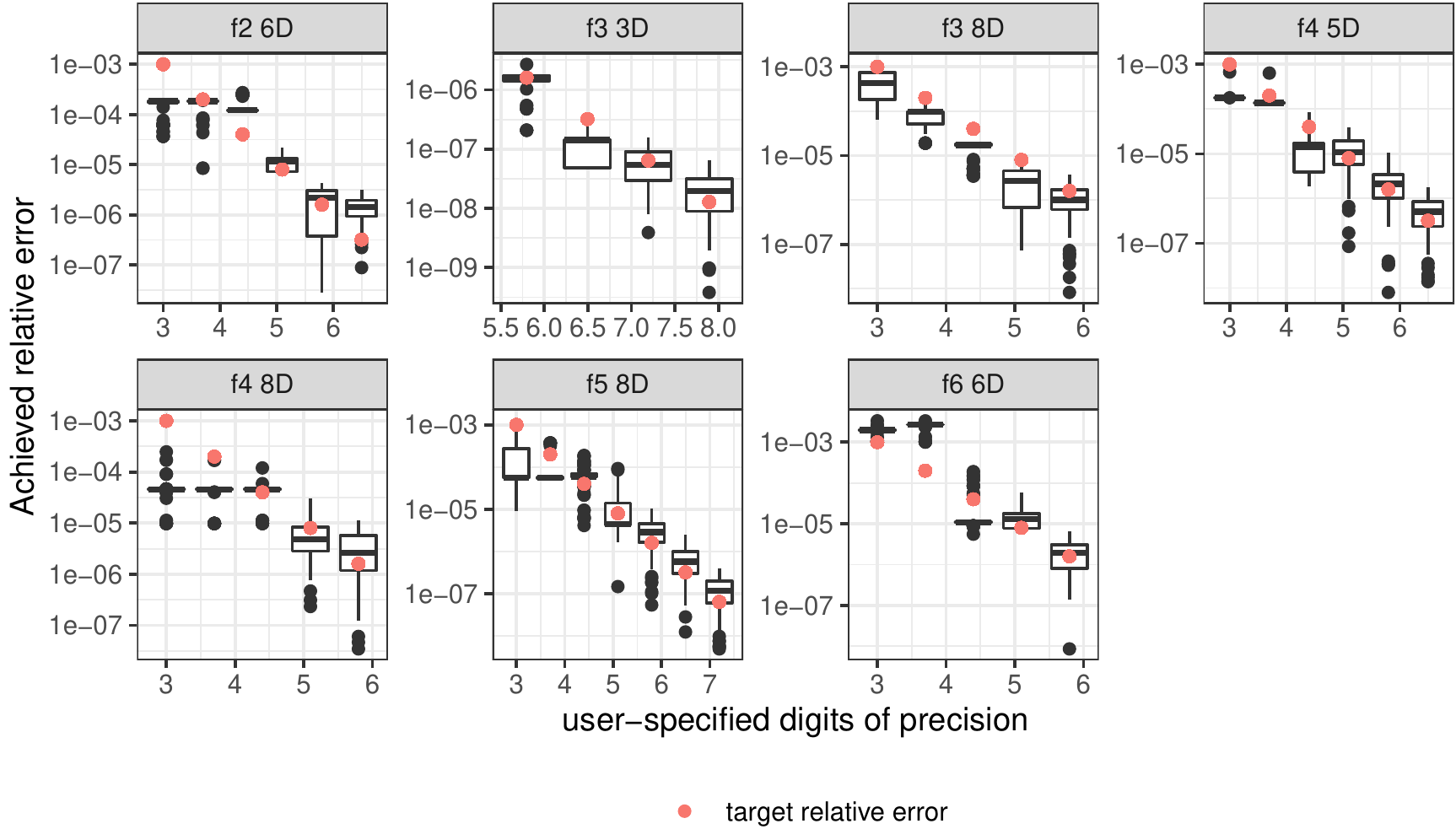}
		\caption{This box plot displays the user-requested relative error tolerance (orange dot) and the achieved relative errors of \textsc{$m$-Cubes} algorithm on the $y$-axis. Each box is a statistical summary of 100 runs. The top and bottom box boundaries indicate the first and third quartiles. The middle line is the value of the median while the vertical lines protruding from the top and bottom box boundaries indicate the minimum and maximum values. The individual points displayed are outliers.}
		\label{fig:boxplot}
	\end{figure}

	We investigate the quality of the \textsc{$m$-Cubes} error-estimates in Figure \ref{fig:boxplot}, where we display multiple 100-run sets of results on each different level of precision for each integrand. The user's requested digits of precision are represented in the $x$-axis, while the true relative error is mapped to the $y$-axis. To make our plots more intuitive, with the $x$-axis representing increasing accuracy requirements, we perform the $-log_{10}(\tau_{rel})$ transformation on the $x$-axis of all plots; this translates ``roughly'' to the required digits-of-precision. We still plot the user's $\tau_{rel}$ for each experiment as the orange point. Since we only plot results for which \textsc{$m$-Cubes} claimed convergence with appropriately small $\chi^{2}$, comparing against the orange point indicates whether the algorithm is as accurate as it claims. 
	
	Due to the randomness of the Monte Carlo samples, there is a wide range of achieved relative error values for the same digits-of-precision. This is to be expected as the error-estimate is interpreted as the standard deviation of the weighted iteration results. Deviation in the results can be more pronounced when generating smaller number of samples which is typical in low-precision runs. In most cases, the number of samples must be increased for higher precisions runs. This leads to a smaller deviation in the results, demonstrated in the figure by the increasingly smaller boxes on the right side of the $x$-axis. This smaller deviation yields improved accuracy, as we observe the box boundaries encompassing the target relative error. We observed similar behavior from \textsc{Gsl}, \textsc{Cuba}, and the \textsc{Vegas} 5.0 Python package on which we performed single-run experiments. 
	
	\subsection{Performance}
	
	\textsc{$m$-Cubes} generates the random numbers and evaluates the integrand within two GPU kernels, \textsc{V-Sample} which additionally stores bin contributions in order to better approximate the distribution of the integrand, and \textsc{V-Sample-No-Adjust} which does not update bin contributions. The execution time of the two kernels, is directly dependent on the number of required function calls per iteration which in turn determines the workload (number of sub-cubes) assigned to each thread. The required number of iterations tends to increase for higher precision runs. For low-precision runs, the same number of samples and iterations can result in convergence. This is why for some integrands ($f_{4,8}$, $f_{5,8}$, $f_{3,3}$, $f_{2,6}$), the three, four, and five digits of precision runs display similar execution time. Missing entries indicate that the corresponding algorithm did not convergence to the required $\tau_{rel}$ in a reasonable amount of time.
	
	We compare \textsc{$m$-Cubes} and gVEGAS by evaluating integrands $1$ to $6$ for various ${\tau}_{rel}$. We observe that \textsc{$m$-Cubes} can be more than one order of magnitude faster. This is attributed to the additional data movement (function evaluations) gVegas requires between CPU and GPU, the smaller number of samples per iteration that are imposed by required memory allocations. By contrast, \textsc{$m$-Cubes} accumulates bin contributions and function evaluations within the GPU and only performs their adjustment on the CPU while the bin boundaries and their contributions are the only data moved between CPU and GPU.
		
	\begin{figure}[hbt]
		\centering
		\includegraphics[width=.95\textwidth]{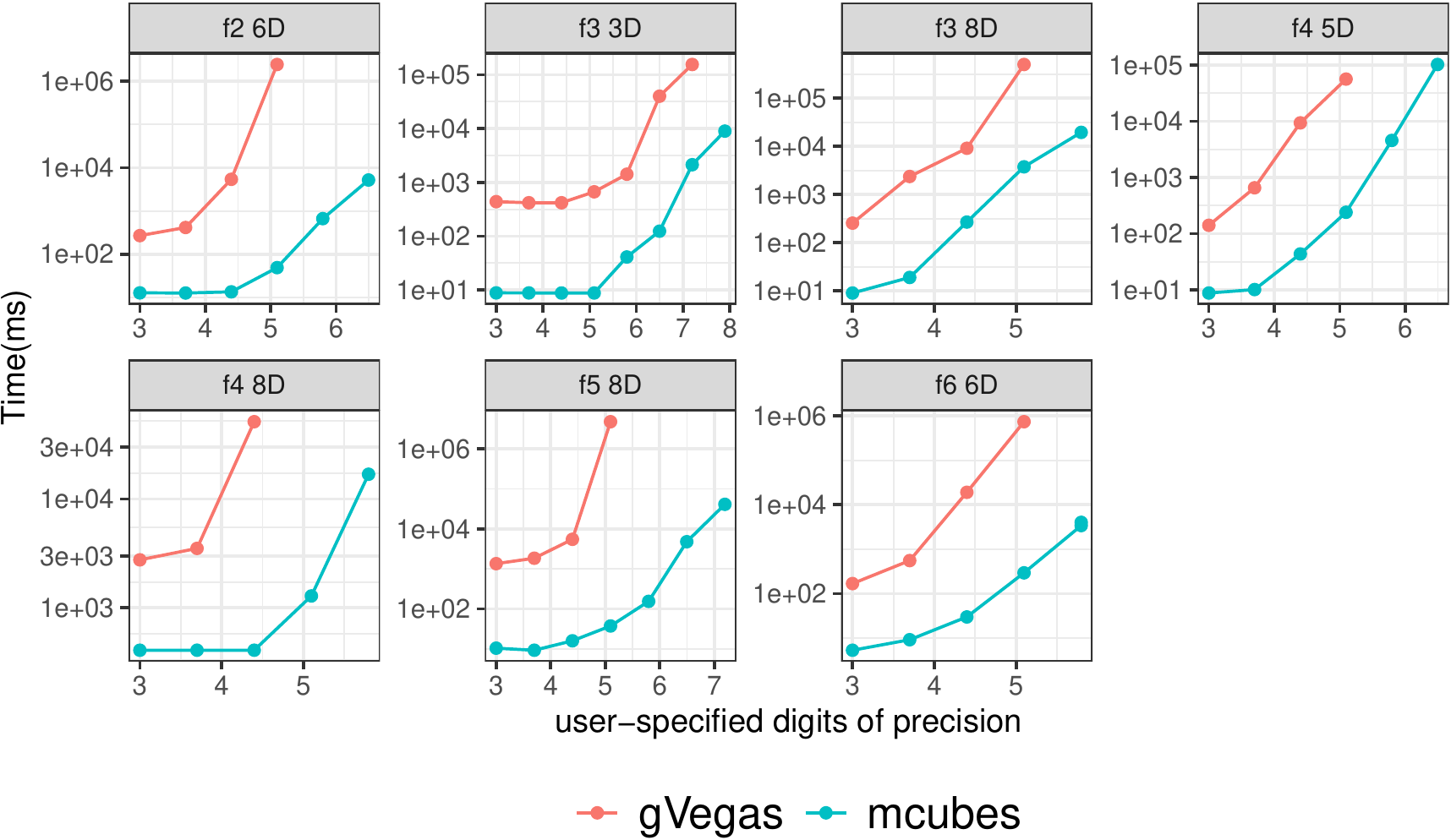}
		\caption{gVegas Comparison}
		\label{fig:gVegasComp}
	\end{figure}
		
	\begin{equation}
		f_A\left(x\right) = \sin\left(\sum_{i=1}^{6} x_i\right)
	\end{equation}
	
	\begin{equation}
		f_B\left(x\right) = \frac{1}{(\sqrt(2\cdot \pi \cdot .01))^9}\exp\left(-\frac{1}{2\cdot (.01)^2} \sum_{i=1}^{9} (x_i)^2\right)
	\end{equation}

	The results of \textsc{ZMCintegral} presented in \cite{WU2020106962} did not include integrands $1$ to $6$ and our experiments showed that \textsc{ZMCintegral} performed slower than serial \textsc{Vegas} in those cases; as we are not aware of the ``best''configuration parameters for those integrands, we do not include such results on the grounds of unfair comparison.	Instead, we use the same parameters on the same integrals reported in \cite{WU2020106962} (integrands  $7$ and $8$). The $f_A$ integrand was evaluated over the range $(0,10)$ on all dimensions, while the integration space of $f_B$ was the range $(-1, 1)$ on all axes. Since ZMCintegral does not accept $\tau_{rel}$ as parameter, we try to match the achieved standard deviation of ZMCintegral for a fair comparison by using a ${\tau}_{rel}$ of $10^{-3}$ and setting the maximum iterations of $10$ and $15$ respectively. We report our results in Table \ref{Tab:ZMCtable}, where we observe a speedup of $45$ and $10$ respectively, though in both cases \textsc{$m$-Cubes} reported significantly smaller error-estimates than ZMCintegral.

	\begin{table}[ht]
		\vspace{-5mm}
		\caption{Comparison with ZMCintegral}
		\centering
		\begin{tabular}{p{2cm}p{2cm}p{2cm} p{2cm}p{2cm}p{1.5cm}}
				\hline
				integrand & alg & true value & estimate & errorest & time (ms) \\ 
				\hline
				$f_A$ & zmc & -49.165073& -48.64740 & 1.98669 & $4.75 \times 10^{4}$ \\ 
				$f_A$ & \textsc{$m$-Cubes} & &-49.27284 & 1.19551 & $1.07 \times 10^{3}$ \\ 
				\hline 
				$f_B$ & zmc & 1.0 & 0.99939 & 0.00133 & $8.30 \times 10^{3}$ \\ 
				$f_B$ & \textsc{$m$-Cubes} &  & 1.00008 & 0.00005 & $9.80 \times 10^{2}$ \\ 
				\hline
		\end{tabular}
		\label{Tab:ZMCtable}
		\vspace{-6mm}
	\end{table}	
		
	\subsection{Cost of function evaluation}	
	
	One of the fundamental operations of \textsc{$m$-Cubes} and all Monte Carlo integration methods is the evaluation of the samples after randomly generating their location in the region space. The execution time for the sample evaluations of the closed-form integrands $1$ to $8$ was typically negligible compared to the total execution time (typically less than $1\%$ and at most $18\%$ in the case of $f_A$). ``Real-world'' integrands can be more costly due to often required non-trivial operations or even expensive memory accesses to look-up tables. In such cases, additional parallelism at the sample evaluation level could provide performance improvement. For example, we could use multiple threads to evaluate a single sample instead of having each thread compute a sample independently. Such operations could involve the parallel generation of the points in each dimensional axis, or even the parallelization of computations requiring multiple look-up operations, such as interpolation, to minimize serial memory accesses.

	\subsection{The \sc{$m$-Cubes1D} variant}
	
	In addition to the \textsc{$m$-Cubes} algorithm, we also provide the variant \textsc{$m$-Cubes1D}. \textsc{$m$-Cubes1D} mirrors \textsc{$m$-Cubes}, with the distinction that the bin boundaries being updated at line 15 in Algorithm 3, are identical on all coordinate axes, thus not requiring the for-loop at line 14. This is beneficial when the integrand $f$ is fully symmetrical, having the same density across each dimension. Thus, one series of atomic additions are required for dimension $j=0$ at line 15. When the bins are then adjusted sequentially after the execution of the \textsc{V-Sample} method, the bins at each dimension will have identical boundaries.
	
\begin{figure}[hbt]
		\centering
			\includegraphics[scale = .4]{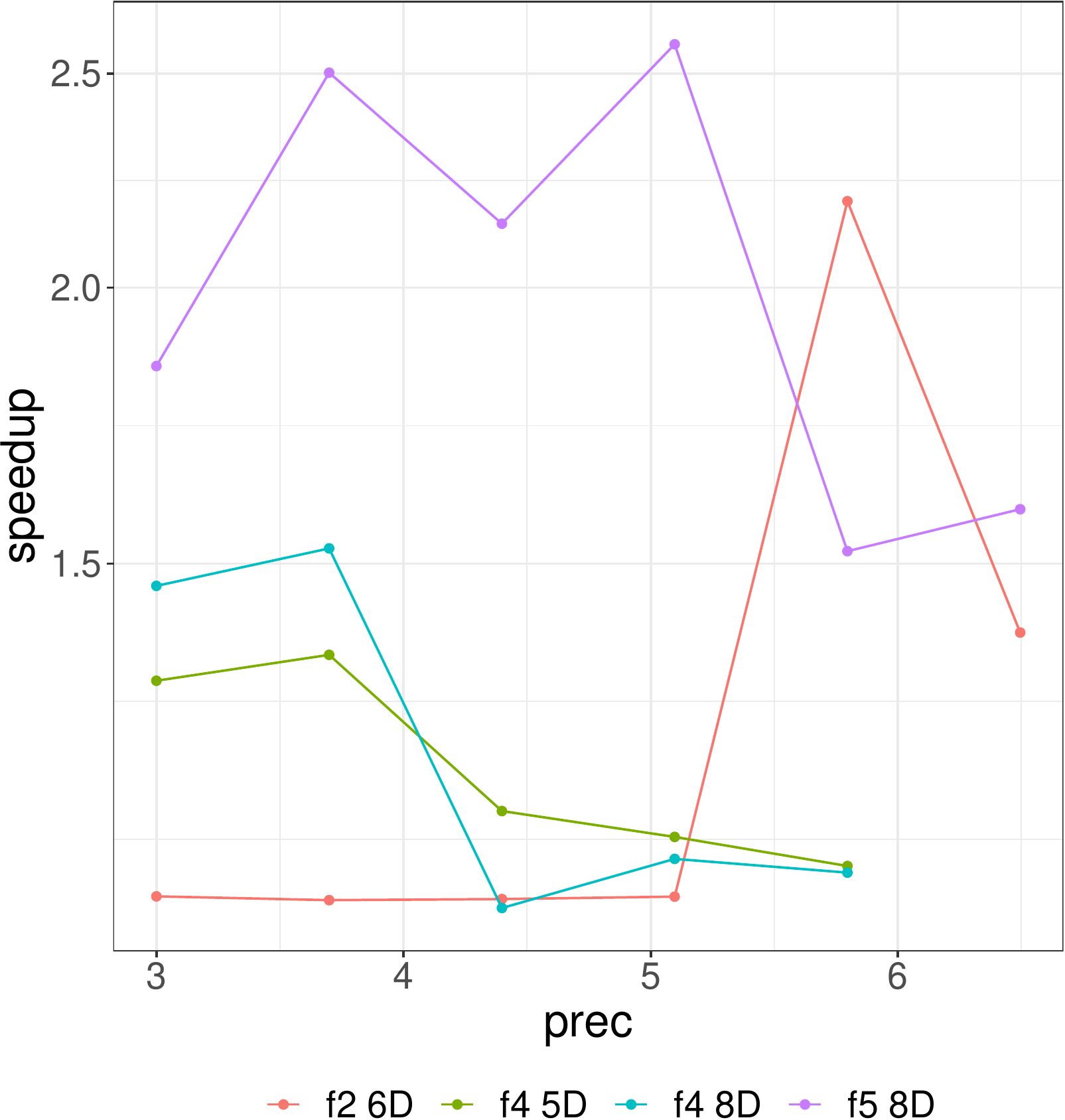}
		\caption{Speedup of \textsc{$m$-Cubes1D} over \textsc{$m$-Cubes} on symmetrical integrands. }
		\label{fig:Mcubes1DOverCPU}
	\end{figure}

	Three of the six integrals presented in section IV, are symmetrical. We performed comparisons between \textsc{$m$-Cubes} and \textsc{$m$-Cubes}1D, which demonstrate a small performance boost in \textsc{$m$-Cubes}1D. In Figure \ref{fig:Mcubes1DOverCPU}, we see speedup of various magnitudes depending on the integrand and degree of precision. Theoretically, both implementations would perform the same bin-adjustments on a symmetrical integrand, and \textsc{$m$-Cubes1D} would require fewer computations for the same effect. We expect that execution of this variant on non-symmetrical integrands, will severely hinder the bin adjustments.
	
\section{Portability}
	
	There are two aspects related to portability: restrictions on the execution platform, and maintaining flexibility when defining an integrand. 
	
	\subsection{Defining integrands in CUDA}
	
	Ideally, an integrator should be ``easy'' to incorporate into existing codes and the integrand definitions should be suitable for execution on various platforms, whether that is CPUs or GPUs regardless of architecture (\textsc{Nvidia}, \textsc{Intel}, \textsc{Amd}, etc.) Additionally, a user should have minimum restrictions when defining the integrand, being allowed to use dynamically created data-structures within the integrand, maintain an integrand state (persistent variables, tabular data, etc.), and define boundaries in the integration space.  
	
	The different memory spaces utilized by a GPU pose a challenge in regards to defining integrands with complex states (non-trivial structures). While the user could potentially interface with \textsc{$m$-Cubes} through the appropriate use of \textsc{Cuda} to handle the different memory spaces, this would severely hinder its ease-of-use and require sufficient knowledge of GPU programming. Additionally, a user who wishes to maintain the option of which platform (CPU, GPU) to execute on, would be forced to write multiple, potentially very different implementations of the same integrand to accommodate the requirements of each platform. To solve this problem, we require the user to define an integrand as a functor to interface with \textsc{$m$-Cubes}. We also supply our own data-structures such as interpolator objects and array-like structures, that handle the GPU related data manipulations internally, but are set and accessed similar to standard library or \textsc{Gsl} equivalents. This allows the user to initialize such objects and structures in a familiar fashion and use them in their defined integrands without having to worry about allocating and transferring data to GPU memory and without having to write any complicated \textsc{Cuda} code. Finally, in regards to ``easily'' using integrators in existing code-bases, \textsc{$m$-Cubes} is implemented as a header-only library.
	
	A use-case demonstrating these features, involves an integrand required for a cosmological study in an astrophysics application. The integrand is six dimensional and requires the utilization of numerous interpolation tables that must be read at run-time and consists of several C++ objects. We evaluated that integrand and compared execution times against the serial \textsc{Vegas} implementation of \textsc{Cuba}. \textsc{$m$-Cubes} returns similar results to those of the \textsc{Cuba} implemenation with appropriate performance. This demonstrates that our solutions pertaining to portability are functional and do not induce any prohibitive costs that would make \textsc{$m$-Cubes} is unsuitable for computationally expensive ``real-world'' integrands.
		
	\subsection{Execution platform portability using kokkos}
		
	We have completed an initial implementation of \textsc{$m$-Cubes} in Kokkos with minimal algorithmic changes to the original \textsc{Cuda} version. The hierarchical parallelism constructs of Kokkos, allow the specification of the same thread-block configuration as required by the \textsc{Cuda} kernels. This makes ``translation'' to Kokkos easy to perform but further optimization is required to maintain performance across architectures. 
	
	We present results on the $f_A$ and $f_B$ integrands in Table \ref{fig:Kokkos}, which displays the kernel time (time executing on GPU) and total time (CPU and GPU time). We evaluated both integrands with the Kokkos version, for three digits of precision on an \textsc{Nvidia} V$100$ GPU. This demonstrates the minimum expected overhead in the range $10$-$15\%$ for the parallel segments of the code, which are expected to cover the majority of execution time. We note that Kokkos can in some cases be faster on the serial code execution. This leads to the low-precision runs on the two integrands being slightly faster in Kokkos. Additional experiments on other integrands show that this is not the case when computational intensity increases. For example, when we compare the running times for the integrand $f_{4,5}$ with $100$ runs for each precision level, Kokkos incurs $20$-$50\%$ overhead.
	
\begin{table}[hbt]
	\vspace{-6mm}
	\caption{Kokkos and \textsc{Cuda} \textsc{$m$-Cubes} Execution Time (ms)}
	\begin{subtable}[h]{0.45\textwidth}
		\centering
		\caption{Execution Time (ms) on $f_A$}
		\begin{tabular}{p{2cm} p{1.5cm}p{1.5cm}}
		\hline
		platform & kernel & total \\ 
		\hline
		\textsc{Cuda} & 829.760 & 1280.318 \\ 
		Kokkos & 968.880 & 1001.035 \\ 
		\hline
		\end{tabular}
		\label{table:kokkos_sinsum_6d}
	\end{subtable}
	\hfill
	\begin{subtable}[h]{0.45\textwidth}
		\centering
		\caption{Execution Time (ms) on $f_B$}
		\begin{tabular}{p{2cm} p{1.5cm}p{1.5cm}}
		\hline
		platform & kernel & total \\ 
		\hline
		\textsc{Cuda} & 664.977 & 1126.529 \\ 
		Kokkos & 726.766 & 767.343 \\ 
		\hline
		\end{tabular}
		\label{table:kokkos_gauss_9d}
	\end{subtable}
	\label{fig:Kokkos}
	\vspace{-5mm}
\end{table}

\section{Conclusion}

We presented \textsc{$m$-Cubes}, a new parallel implementation of the widely used \textsc{Vegas} multi-dimensional numerical integration algorithm for execution on GPUs. $m$-Cubes is a portable header-only library, with a modern interface and features that allow easy interfacing and requires no knowledge of GPU programming to use. We also supply infrastructure to facilitate the definition of complex and stateful integrands. Our experiments on a standard set of challenging integrals and a complex stateful integrand consisted of numerous C++ objects, demonstrate superior performance over existing GPU implementations. Furthermore, We supply the variant \textsc{$m$-Cubes1D} to accelerate evaluation of symmetrical integrals. We also provide an initial Kokkos implementation to allow execution on non-\textsc{Nvidia} GPUs.

\bibliography{refs}

\begin{thebibliography}{10}
\providecommand{\url}[1]{\texttt{#1}}
\providecommand{\urlprefix}{URL }
\providecommand{\doi}[1]{https://doi.org/#1}

\bibitem{gvegascp}
 (2017), \url{https://github.com/lbiedma/gVegascp}

\bibitem{madgraph}
 (2020), \url{https://xgitlab.cels.anl.gov/whopkins/MadgraphGPU}

\bibitem{qmcGPU}
Borowka, S., Heinrich, G., Jahn, S., Jones, S., Kerner, M., Schlenk, J.: A gpu
  compatible quasi-monte carlo integrator interfaced to pysecdec. Computer
  Physics Communications  \textbf{240},  120–137 (Jul 2019).
  \doi{10.1016/j.cpc.2019.02.015},
  \url{http://dx.doi.org/10.1016/j.cpc.2019.02.015}

\bibitem{Carrazza}
Carrazza, S., Cruz-Martinez, J.M.: {VegasFlow: accelerating Monte Carlo
  simulation across multiple hardware platforms}. Comput. Phys. Commun.
  \textbf{254},  107376 (2020). \doi{10.1016/j.cpc.2020.107376}

\bibitem{vegasflow_package}
Cruz-Martinez, J., Carrazza, S.: N3pdf/vegasflow: vegasflow v1.0 (Feb 2020).
  \doi{10.5281/zenodo.3691926}, \url{https://doi.org/10.5281/zenodo.3691926}

\bibitem{CarterEdwards20143202}
Edwards, H.C., Trott, C.R., Sunderland, D.: Kokkos: Enabling manycore
  performance portability through polymorphic memory access patterns. Journal
  of Parallel and Distributed Computing  \textbf{74}(12),  3202 -- 3216 (2014).
  \doi{https://doi.org/10.1016/j.jpdc.2014.07.003},
  \url{http://www.sciencedirect.com/science/article/pii/S0743731514001257},
  domain-Specific Languages and High-Level Frameworks for High-Performance
  Computing

\bibitem{GenzTest}
Genz, A.: Testing multidimensional integration routines. In: Proc. of
  International Conference on Tools, Methods and Languages for Scientific and
  Engineering Computation. p. 81–94. Elsevier North-Holland, Inc., USA (1984)

\bibitem{Goda2019RecentAI}
Goda, T., Suzuki, K.: Recent advances in higher order quasi-monte carlo
  methods. arXiv: Numerical Analysis  (2019)

\bibitem{KanzakiJ2011MCio}
Kanzaki, J.: Monte carlo integration on gpu. The European physical journal. C,
  Particles and fields  \textbf{71}(2), ~1--7 (2011)

\bibitem{LEPAGE2021110386}
Lepage, G.P.: Adaptive multidimensional integration: vegas enhanced. Journal of
  Computational Physics  \textbf{439},  110386 (2021).
  \doi{https://doi.org/10.1016/j.jcp.2021.110386},
  \url{https://www.sciencedirect.com/science/article/pii/S0021999121002813}

\bibitem{PETERLEPAGE1978192}
{Peter Lepage}, G.: A new algorithm for adaptive multidimensional integration.
  Journal of Computational Physics  \textbf{27}(2),  192--203 (1978).
  \doi{https://doi.org/10.1016/0021-9991(78)90004-9},
  \url{https://www.sciencedirect.com/science/article/pii/0021999178900049}

\bibitem{pagani}
Sakiotis, I., Arumugam, K., Paterno, M., Ranjan, D., Terzi\'{c}, B., Zubair,
  M.: PAGANI: A Parallel Adaptive GPU Algorithm for Numerical Integration.
  Association for Computing Machinery, New York, NY, USA (2021),
  \url{https://doi.org/10.1145/3458817.3476198}

\bibitem{9485033}
Trott, C.R., Lebrun-Grandié, D., Arndt, D., Ciesko, J., Dang, V., Ellingwood,
  N., Gayatri, R., Harvey, E., Hollman, D.S., Ibanez, D., Liber, N., Madsen,
  J., Miles, J., Poliakoff, D., Powell, A., Rajamanickam, S., Simberg, M.,
  Sunderland, D., Turcksin, B., Wilke, J.: Kokkos 3: Programming model
  extensions for the exascale era. IEEE Transactions on Parallel and
  Distributed Systems  \textbf{33}(4),  805--817 (2022).
  \doi{10.1109/TPDS.2021.3097283}

\bibitem{WU2020106962}
Wu, H.Z., Zhang, J.J., Pang, L.G., Wang, Q.: Zmcintegral: A package for
  multi-dimensional monte carlo integration on multi-gpus. Computer Physics
  Communications  \textbf{248},  106962 (2020).
  \doi{https://doi.org/10.1016/j.cpc.2019.106962},
  \url{https://www.sciencedirect.com/science/article/pii/S0010465519303121}

\end{thebibliography}
\end{document}